\documentstyle[12pt]{article}
\begin{document}


\begin{center}
 {\Large \bf  A q-deformed Version of the Heavenly Equations}
\vskip 2truecm
 J. F. Pleba\'nski  and H. Garc\'{\i}a-Compe\'an \\
{\it Departamento de F\'\i sica}\\
 {\it Centro de Investigaci\'on y de Estudios Avanzados del I.P.N.} \\
{\it Apdo. Postal 14-740, 07000, M\'exico, D.F., M\'exico.} \\
 \vskip2.6cm
 {\Large \bf Abstract}
\end{center}

\noindent
 Using a $q$-deformed  Moyal algebra associated with the group of
 area preserving diffeomorphisms of th two-dimensional torus $T^2$,
 sdiff$_q (T^2)$, a $q$-deformed version
 for the Heavenly equations is given. Finally, the two-dimensional chiral
 version of Self-dual gravity in this $q$-deformed context is briefly
 discussed.

\vskip1.5truecm


\vfill \eject
\baselineskip 24pt

\section{Introduction}

  New directions in mathematical physics seem to converge to Self-dual
 gravity in different ways. A most interesting formulation is due to
 Pleba\'nski [1]. In this Ref. it is found that the Self-dual gravity is
described
 by the non-linear second order partial differential equation for a
 holomorphic function (the Heavenly equation).

 At present time there exists great interest to find a deep relation
 between Self-dual Yang-Mills gauge theory and Self-dual gravity in
 dimension four [2-5]. In these papers Self-dual gravity is obtained
 from a dimensional reduction of the Self-dual Yang-Mills equations on a
 4-dimensional flat space-time ${\cal M}$ with infinite dimensional gauge
 group SU$(\infty)$.
 This group was taken to be the group of area preserving diffeomorphisms
 of the two-surface $\Sigma$, SDiff$(\Sigma)$.

 Some symmetries involved here are infinite dimensional. In particular the
 large-$N$ limit of the algebra of Zamolodchikov $W_N$, i.e. $W_\infty$, plays
 a crucial role. \ The connection between $W_\infty$ algebras and Self-dual
 gravity was pointed out for the first time by Bakas in the first paper of
 Ref [6]. \ In the present paper we use this fact.

 In a similar spirit, in the paper [7], we show that working with the
 Self-dual Yang-Mills
 fields on the 4-dimensional flat manifold ${\cal M}$ with the Lie algebra
 sdiff$(\Sigma)$-valued connection 1-form, the first and the second Heavenly
 equations emerge in a natural manner. Here
 sdiff$(\Sigma)$ is the Lie algebra of SDiff$(\Sigma)$.

 On the other hand, there was a considerable interest recently in the
 application of
 quantum groups and noncommutative geometry in gauge theory [8,9,10]. In
 Refs. [11,12] the Yang-Mills gauge theory on the classical
 space-time was defined using the quantum group SU$_q(2)$ as the ``symmetry
 group''. In Ref. [13] it is shown how General Relativity can be put in the
 context of noncommutative
 geometry. Until now a general theory
 for Einstein's theory of gravity including quantum groups as
 well as quantum spaces has not yet been considered. This work forms a
 part of our research for the case of Self-dual gravity. We think that
 this generalization
 may include some new insights in a realistic theory of quantum gravity.

 In this paper we use a $q$-deformed version of the Moyal algebra
 associated with the Lie algebra of the group
 of area preserving diffeomorphisms of the torus $T^2 \ \ (\Sigma = T^2),$
 sdiff$_q (T^2)$ [14, 15] in order to show new insights in four-dimensional
 Self-dual gravity. Specifically, we obtain a $q$-deformed version of the
 first
 and the second heavenly equations of Self-dual gravity [1]. This forms
 part
 of our attempts to construct the general theory of ${\cal H}$ and ${\cal
 H}{\cal H}$
 quantum spaces with quantum group as the symmetry group in a close philosophy
 with [16].

 The paper is organized as follow, in section 2 we describe the necessary tools
 and the quantum algebras used in section 3.  The section  3 is devoted to
obtain
 a $q$-deformed version for the Heavenly equations. In section 4 we briefly
 discuss the recent two-dimensional chiral formulation of Self-dual
 gravity given by Husain in Ref. [17] in the context of $q$-deformed
 algebras. Finally, in section 5, we give our conclusions.

\vskip 2truecm
\section{\bf Basic Tools and Quantum Algebras}

 The large $N$-limit of the algebra $W_N$ generated by the primary conformal
 fields with spins $1,2,...,N$ is known as a $W_{\infty}$ algebra. In
 Ref. [18] Bakas found that this algebra provides the representation of an
 infinite dimensional (sub)algebra of the area preserving diffeomorphisms
 of the plane.

 Thus, taking the generators $W_{(k,m)}$ (of spin $k, k \geq 0, m \in
 {\bf Z})$ of the $W_\infty$ algebra, they satisfy the relation

\begin{equation}
 [W_{(k,m)}, W_{(l,n)}] = \{(l+1) (n+1) - (k+1) (m+1) \} W_{(k+l, m+n)}.
\end{equation}

 As is well known, if one takes the large $N$-limit of the Lie algebra su$(N)$
it can be identified in a natural manner with
 sdiff$(T^2)$, {\it i.e.} su$(\infty) \simeq$ sdiff$(T^2)$ [19].

 The algebra sdiff$(T^2)$come given by the Poisson algebra

\begin{equation}
 [L_f, L_g] = L_{\{f, g\}}
\end{equation}

\noindent
 where $\{ f,g\}$ is the usual Poisson bracket and $L_f = \frac{\partial f}{
 \partial x} \ \frac{\partial }{\partial y} - \frac{\partial f}{\partial y}
 \frac{\partial} {\partial x}.$

 Taking a basis for the torus

\begin{equation}
 f \to f_{\bf m} =\mbox{exp} ({\bf im}\cdot {\bf x}), \ \ \ g\to f_{\bf n}
 = \mbox{exp} ({\bf in} \cdot {\bf x}),
\end{equation}

\noindent
 where ${\bf m}$ and ${\bf n}$ are 2-vectors with integer entries,
 ${\bf m} = (m_1, m_2)$ and
 ${\bf x} = (x_1, x_2)$. In this basis the Poisson algebras becomes the algebra
sdiff$(T^2)$

\begin{equation}
 [L_{\bf m}, L_{\bf n}] = ({\bf m} \times {\bf n}) L_{{\bf m}+{\bf n}}
\end{equation}

\noindent
 where ${\bf m} \times {\bf n} = m_1 n_2 - m_2 n_1.$

 A deformation of the Poisson algebra (2) is the Moyal algebra.  For the torus,
 the Moyal algebra reads

\begin{equation}
 [L_{\bf m}, L_{\bf n}] = \frac{1}{\kappa} Sin ({\kappa} {\bf m} \times {\bf
n})    L_{{\bf m}+{\bf n}}
\end{equation}

\noindent
$\kappa$ is here the parameter of the deformation.

 A $q$-deformed version of this $W_\infty$ algebra is

 \[ [W^q_{(k,m)}, W^q_{(l,n)}]_q = W^q_{(k,m)} W^q_{(l,n)} - q \cdot
 W^q_{(l,n)} W^q_{(k,m)}  \]

\begin{equation}
 = \{ (l+1) (n+1) - (k+1) (m + 1)\} W^q_{(k+l,m+n)}
\end{equation}

\noindent
 where $q$ is a complex parameter, in particular a root of unity, $q =
\mbox{exp} (ih),$ where $h$ is a real number.

 This algebra corresponds to a $q$-deformed version of sdiff$(\Sigma)$,
 namely sdiff$_q\Sigma$.  \ However it is not very general because it
 is limited to anti-commute generators\footnote{ We thank Prof. C. Zachos for
pointing out this consequence.}.

 The Moyal algebra (5) is also a Lie algebra and therefore can be $q$-deformed.
 There exist a $q$-deformation of the Moyal bracket (`quantum Moyal') proposed
 for the first time by Devchand, Fairlie, Fletcher and Sudbery [15].

 On the other hand, there is other possible $q$-deformation of sdiff$(\Sigma)$.
 In Ref. [14], Fairlie found such a $q$-deformation from the construction of
 algebras of $q$-symmetrized polynomials in the $q$-Heisenberg operators $P$
 and $Q$. That is, operators which satisfy the $q$-deformed Heisenberg
 algebra

 \[ PQ - q QP = i \lambda. \]

\noindent
 where $\lambda$ is a real parameter. Fairlie shown [14] that the $q$-deformed
 Heisenberg algebra leads to a $q$-deformation of the Moyal algebra

\begin{equation}
 q^{{\bf n}\times {\bf m}} \cdot W_{\bf m} W_{\bf n} - q^{{\bf m} \times
 {\bf n}} \cdot W_{\bf n} W_{\bf m} = (\omega^{{\bf m}\times {\bf n}/2}
 - \omega^{{\bf n}\times {\bf m}/2}) W_{{\bf m}+{\bf n}} + {\bf a} \cdot {\bf
m}
 \delta_{{\bf m}+{\bf n},0}
\end{equation}

\noindent
 where ${\bf a}$ is a constant 2-vector which characterizes the central
extension
 (see also Ref. [20]). The classical limit of (7) gives precisely the Moyal
 algebra (5).

 In this algebra (7), appears two parameters $q$ and $\omega$.  If it is taked
 $\omega = $ exp $(i {\kappa})$ and $q = $ exp $(ih) \to 1 (\mbox {or} \ h \to
0)$ we recover of course the Moyal algebra (5).

 In order to apply the above $q$-deformed Moyal algebra is convenient to take
 ${\bf m} = (m_1, 0), {\bf n} = (0, n_2)$ in Eq. (7). Thus we have

\begin{equation}
 q^x \cdot W_{\bf m} W_{\bf n} - q^{-x} \cdot W_{\bf n} W_{\bf m} =
 (\omega^{x/2} - \omega^{-x/2} )
 W_{{\bf m}+{\bf n}} + a_1 m_1 \delta_{(m_1, n_2),0}
\end{equation}
where $x \equiv m_1 n_2$.

 We will apply the $q$-deformed Moyal  algebra to the compatibility conditions
for the usual
 Self-dual Yang-Mills equations on ${\cal M}$  (``classical'' space) with
 local coordinates $\{t,x,y,z\}$ and after this we compare the corresponding
 results.

 Redefining the local coordinates on ${\cal M}$ to be, $ \ \alpha
 = t + iz,
 \ \ \bar \alpha =t - i z, \ \ \beta= x + i y$ and $\bar \beta=x - iy$
 the Self-dual Yang-Mills equations are

\begin{equation}
 F_{\alpha \beta} = 0, \ \ F_{\bar \alpha \bar \beta} = 0, \ \
 F_{\alpha \bar \alpha} + F_{\beta \bar \beta} = 0
\end{equation}

\noindent
 where

\begin{equation}
 F_{ij} = \partial_iA_j - \partial_j A_i + [A_i,A_j]
\end{equation}

\noindent
 and $i,j \in \{\alpha,\bar \alpha, \beta,\bar \beta\}$. The compatibility
 condition reads

 \[ [\partial_{\alpha} + \lambda \partial_{\bar \beta},A_{\beta}- \lambda
 A_{\bar \alpha}] - [\partial_{\beta} - \lambda \partial_{\bar \alpha},
 A_{\alpha} + \lambda A_{\bar \beta}] \]

\begin{equation}
 = [A_{\alpha} + \lambda A_{\bar \beta}, A_{\beta} - \lambda A_{\bar
 \alpha}].
\end{equation}

  Now, we will work with the bundle

\begin{equation}
  {\rm SDiff}_q (\Sigma) \ \ - \ \ P \buildrel \rm \pi \over {\rightarrow}
  {\cal M}
\end{equation}

\noindent
 with the connection 1-form $A_i$ on ${\cal M}$ taking values precisely on
 sdiff$_q (T^2)$, {\it i.e.}

\begin{equation}
 A_i= \Phi_ {i,s} {\partial \over \partial r} - \Phi_{i,r}
 {\partial \over \partial s} \in {\rm sdiff}_q(T^2)
\end{equation}

\noindent
 where $\Phi = \Phi (\alpha,\bar \alpha, \beta,\bar \beta;q)$ and $r,s$ are
 local coordinates on the two-dimensional torus $T^2$.
 They are the generators of sdiff$_q(T^2)$ and satisfy the algebra (7) or (8).

 The compatibility condition in the $q$-deformed version is

\[
 [\partial_{\alpha} + \lambda \partial_{\bar \beta},A_{\beta}- \lambda
 A_{\bar \alpha}] -[\partial_{\beta} - \lambda \partial_{\bar \alpha},
 A_{\alpha} + \lambda A_{\bar \beta}]
\]

\begin{equation}
  = [A_{\alpha} + \lambda A_{\bar \beta}, A_{\beta} - \lambda A_{\bar
 \alpha}]_q.
\end{equation}

 This equation will provide us the way of obtaining the $q$-deformed
 version of Heavenly equations.

\vskip 2truecm

\section{ \bf The q-deformed Heavenly Equations}

 We start by substituting  Eq. (13) into (14) and use the $q$-deformed
 commutation relations (8). Comparing powers of $\lambda$ at order zero,
 two and one respectively, we obtain the set of equations

\[
 \Phi_{\alpha,\beta s} - \Phi_{\beta, \alpha s} + (q^x \cdot \Phi_{\alpha, r}
 \Phi_{\beta, ss} + q^{-x} \cdot \Phi_{\beta, s} \Phi_{\alpha, rs}) - (q^x
\cdot \Phi_{\alpha, s} \Phi_{\beta, rs} + q^{-x} \cdot \Phi_{\beta, r}
\Phi_{\alpha, ss})
\]

\begin{equation}
 + {\cal F} (s,\alpha, \bar \alpha, \beta, \bar \beta,q)=0,
\end{equation}

\[
 \Phi_{\bar \alpha, \bar \beta s} - \Phi_{\bar \beta, \bar \alpha s} +
 (q^x \cdot \Phi_{\bar \alpha, r}
 \Phi_{\bar \beta, ss} + q^{-x} \cdot \Phi_{\bar \beta, s} \Phi_{\bar \alpha,
 rs}) - (q^x \cdot \Phi_{\bar \alpha,
 s} \Phi_{\bar \beta, rs} + q^{-x} \cdot \Phi_{\bar \beta, r} \Phi_{\bar
\alpha,ss})
\]

\begin{equation}
 + \bar{\cal F} \ (s,\alpha, \bar \alpha, \beta, \bar \beta,q) = 0,
\end{equation}

\[
 \Phi_{\alpha, \bar \alpha s} - \Phi_{\bar \alpha, \alpha s} + \Phi_{\beta,
 \bar \beta s} - \Phi_{\bar \beta, \beta s} + [(q^{-x} \cdot \Phi_{\bar \alpha,
s}\Phi_{\alpha, rs} + q^x \cdot \Phi_{\alpha, r} \Phi_{\bar \alpha, ss}) -
 (q^{-x} \cdot \Phi_{\bar \alpha, r} \Phi_{\alpha, ss}
\]

\[
 + q^x \cdot \Phi_{\alpha, s}
 \Phi_{\bar \alpha, rs}) + (q^{-x} \cdot \Phi_{\beta, r} \Phi_{\bar \beta, ss}
+ q^x \cdot
 \Phi_{\bar \beta, s} \Phi_{\beta, rs}) - (q^{-x} \cdot \Phi_{\beta, s}
\Phi_{\bar \beta,
 rs} + q^x \cdot \Phi_{\bar \beta, r} \Phi_{\beta, ss})]
\]

\begin{equation}
+ {\cal G} \ (s,\alpha, \bar \alpha, \beta, \bar \beta,q) = 0,
\end{equation}

\noindent
 here $\Phi_{\alpha,rs} = {\partial^2 \Phi_{\alpha} \over \partial r
 \partial s}$, etc.

\vskip 1truecm
\noindent

\begin{center}
 {\bf  A. \ \ The q-deformed First Heavenly Equation}
\end{center}

 Now, assuming

\[
 \Phi_{\alpha} = \Omega_{,\alpha},  \ \ \Phi_{\beta} =
 \Omega_{,\beta}, \ \  \Phi_{\bar \alpha} = \Phi_{\bar \beta} = 0,
\]

\begin{equation}
 \bar{\cal F} = 0 = {\cal G}, \ \    {\cal F} = {\cal F}(s, \alpha,\beta, q), \
\   {\cal F} \not= \bar{\cal F} \
\end{equation}

\noindent
 being $\Omega = \Omega(\alpha, \beta, r,s;q)$ some holomorphic function of
 its arguments. Thus, Eqs. (16) and (17) are satisfied trivially and
 Eq. (15) gives

\begin{equation}
 q^x \cdot \Omega_{,\alpha r} \Omega_{,\beta ss} + q^{-x} \cdot \Omega_{,\beta
s}
 \Omega_{,\alpha rs} - (q^x \cdot \Omega_{,\alpha s}\Omega_{,\beta rs} + q^{-x}
\cdot
 \Omega_{,\beta r} \Omega_{,\alpha ss}) + {\cal F} (s, \alpha, \beta , q) = 0
\end{equation}

\noindent
 where $\Omega_{,\alpha r s}= \frac{\partial^3 \Omega}{\partial \alpha \partial
 r \partial s}$ etc. After the change of coordinates $r {\cal F} \to r$, this
 equation leads directly to

\begin{equation}
 q^x \cdot \Omega,_{\alpha r} \Omega,_{\beta ss} +
 q^{-x} \cdot \Omega,_{\beta s} \Omega,_{\alpha r s} - (q^x \cdot
 \Omega,_{\alpha s} \Omega_{, \beta r s} + q^{-x} \cdot \Omega, _{\beta r}
\Omega_{, \alpha s s} = 1.
\end{equation}

 Notice that the $q$-deformed first Heavenly
 equation is a {\it third} order non-linear partial differential equation
 (TONLPDE).

 When the parameter $q \rightarrow 1$ (or equivalently $h \rightarrow
 0$), we recover the first heavenly equation in its usual form [1]

\[
 \Omega_{,\alpha r} \Omega_{,\beta s} - \Omega_{,\alpha s}
 \Omega_{,\beta r} = 1.
\]

\vskip 1truecm
\begin{center}
 {\bf B. \ \ The q-deformed Second Heavenly Equation}
\end{center}
 Taking

\[
 \Phi_{\alpha} = \theta_{,s}, \ \  \Phi_{\beta} = - \theta_r,
 \ \   \Phi_{\bar \alpha} = \Phi_{\bar \beta} = 0,
\]

\begin{equation}
 \bar{\cal F} = 0 = {\cal G}, \ \  {\cal F} = {\cal F}(s,\alpha,
 \beta,q), \ \   \theta = \theta(\alpha,\beta,r,s;q), \ \ {\cal F} \not=
 \bar{\cal F}.
\end{equation}

\noindent
 Thus, Eqs. (16) and (17) hold trivially and Eq. (15) yields

 \begin{equation}
 q^x \cdot \theta_{,ss}\theta_{,rrs} + q^{-x} \cdot \theta_{,rr} \theta_{,sss}
 - (q^x +q^{-x} )
 \theta_{,rs} \theta_{,rss} + \theta_{,r \alpha s} + \theta_{,s \beta s} +
 {\cal F} (s,\alpha,\beta,q) = 0.
\end{equation}

 Now, making the substitution

\begin{equation}
 \Theta = \theta + r s f, \ \ \ \ \ \   f= f(\alpha,\beta,q), \ \
 f_{,\alpha} = {\cal F}
\end{equation}

\noindent
 (with $\Theta = \Theta(\alpha,\beta,r,s;q))$ into Eq. (22) we finally obtain

\begin{equation}
 (q^x \cdot \Theta_{,ss}\Theta_{,rrs} + q^{-x} \cdot \Theta_{,rr}
\Theta_{,sss})  - (q^x + q^{-x} )
 \Theta_{,rs} \Theta_{,rss} + \Theta_{,r \alpha s} + \Theta_{,s \beta s} = 0.
\end{equation}

 This equation is again a TONLPDE.
 Taking the limit $q\rightarrow 1$ we recover of course the second
 Heavenly equation [1]

 \[
 \Theta_{,rr}\Theta_{,ss} - \Theta^2_{,rs} + \Theta_{,r \alpha} +
 \Theta_{,s \beta} = 0.
 \]

\vskip 2truecm

\section{\bf  Comments on the q-deformed Chiral Model
 of the Self-dual Gravity }

 In this last section
 we make some comments about the possible generalization
 of the Husein work [17]. In this paper it is shown that Self-dual
 Einstein equations are equivalent to the two-dimensional Chiral Model
 with gauge group, precisely SDiff$(\Sigma)$.

 Since the gauge group here is SDiff$(\Sigma)$ the connection 1-form are
 sdiff$(\Sigma)$-valued. Thus, we will generalize some results in [17] using
 the $q$-deformed Moyal algebra (8) associated with sdiff$_q(T^2)$ instead
 of the Moyal algebra (5) associated with sdiff$(T^2)$.

 The Ashtekar-Jacobson-Smolin formulation of Self-dual gravity [21] leads
 to a set of equations on the four manifold ${\cal M}^4 = {\cal K}^3 \times
 R$ with local complex coordinates $\{x_0,x_1,x_2,$ $x_3\}$.  The equations
 are (see also [22])

\begin{equation}
 {\rm Div} V_i^a = 0
\end{equation}

\begin{equation}
 {\partial V_i^a \over \partial t} = {1\over 2} \epsilon_{ijk}
 [V_i,V_k]^a
\end{equation}

\noindent
 where $V_i^a$ are three spatial vector fields on ${\cal K}^3$ and $[,]$
 is the Lie bracket. The solutions of these equations lead to the
 self-dual metric

\begin{equation}
 g^{ab} = (detV)^{-1}[V_i^a V_j^b \delta^{ij} + V_0^aV_0^b]
\end{equation}

\noindent
 where $i,j=1,2,3$ and $V_0^a$ is a vector field used in the decomposition
 ${\cal M}^4 = {\cal K}^3 \times {\cal R}$.  The Eq. (26) is equivalent to the
equations

\begin{equation}
 [{\cal T},{\cal X}] = [{\cal U},{\cal V}] = 0,
\end{equation}

\begin{equation}
 [{\cal T},{\cal U}] + [{\cal X},{\cal V}] = 0,
\end{equation}

\noindent
 where ${\cal T}=V_0 + iV_1$, ${\cal X} = V_2 - iV_3$, ${\cal U} = V_0 -
 iV_1$ and ${\cal V} = V_2 + iV_3$. Using the gauge freedom we can put
 in terms of the new coordinates $\beta = x_0 + i x_1$, $\alpha = x_2 - i
 x_3$, $u = x_0 - i x_1$ and $v = x_2 + i x_3$

\begin{equation}
 {\cal T}^a = ({\partial \over \partial \beta})^a, \ \ \ \ \ \ \  {\cal X}^a
 = ({\partial \over \partial \alpha})^a.
\end{equation}

 As Husein showed [17], the Eqs. (28) and (29) are equivalent to the
 two-dimensional Chiral Model on the plane with coordinates $(\alpha , \beta)$.

 The relation with the first Heavenly equation arises in this context when we
 take

\begin{equation}
 {\cal U} = -\Omega ,_{\alpha s} {\partial \over \partial r} +
 \Omega ,_{\alpha r}{\partial \over \partial s},
\end{equation}

\begin{equation}
 {\cal V} = \Omega ,_{\beta s} {\partial \over \partial r} -
 \Omega ,_{\beta r}{\partial \over \partial s}.
\end{equation}

\noindent
 These are precisely the generators of the Lie algebra sdiff$(T^2)$
 ({\it i.e.} ${\cal U,V} \in$ sdiff$(T^2)$) just as
 Eq. (13). As we saw before these generators satisfy the Poisson algebra given
by (2). The Eq. (28) leads directly to the
 first Heavenly equation in the form

 \[
 \Omega_{,\alpha r}\Omega_{,\beta s} - \Omega_{, \alpha s}\Omega_{,\beta r} =
1.
 \]

 Thus, the natural question arises:
 what is the modification of the first Heavenly equation when ${\cal U,V}$
 are sdiff$_q(\Sigma)$-valued? To see this, notice that the only
 modification in the field equations is

\begin{equation}
 [{\cal U},{\cal V}]_q = q^x \cdot {\cal U}{\cal V} - q^{-x} \cdot
 {\cal V}{\cal U}
\end{equation}

\noindent
 where now, ${\cal U}$,${\cal V}$ depends on the parameter $q$ and $\Omega =
 \Omega(\alpha , \beta ,r,s;q)$. Substituting (31) and (32) into (33) we obtain
after a few  steps

\begin{equation}
 q^x \cdot
 \Omega_{,\alpha r} \Omega_{,\beta ss} + q^{-x} \cdot \Omega_{,\beta s}
 \Omega_{,\alpha rs} - (q^x \cdot \Omega_{,\alpha s}\Omega_{,\beta rs} +
 q^{-x} \cdot
 \Omega_{,\beta r} \Omega_{,\alpha ss}) = {\cal O}(s,\alpha , \beta ;q)
\end{equation}

\noindent
 where ${\cal O}$ is some holomorphic function its argument. Taking ${\cal O}
(s, \alpha ,\beta ; q) = 1$ we obtain the $q$-deformed first Heavenly equation
from the $q$-deformed two dimensional Chiral Model . It is

\begin{equation}
 q^{x} \cdot \Omega_{,\alpha r} \Omega_{,\beta ss} + q^{-x} \cdot
\Omega_{,\beta s}
 \Omega_{,\alpha rs} - (q^{x} \cdot \Omega_{,\alpha s}\Omega_{,\beta rs} +
q^{-x} \cdot \Omega_{,\beta r} \Omega_{,\alpha ss}) = 1
\end{equation}

\noindent
 and corresponds exactly to the result given in Eq. (20)  for the Self-dual
 Yang-Mills case.  \ This equation is the
 $q$-deformed version of the Eq. (29) of the first Ref. [17]. Finally,
 taking the $q \rightarrow 1$ we obtain again the first heavenly
 equation in its usual form.

\vskip 2truecm

\section {\bf Conclusions }

 In this paper we have considered some applications of the $q$-deformed
 Moyal algebra (8) associated with the group  of area preserving
 diffeomorphisms of the torus
 $T^2$. Since the Heavenly equations are very closely related  with
 this symmetry [23,24], a $q$-deformed version of these equations is a
 generalization for they. In
 particular, we showed that the modification prevents the terms in
 Eqs. (20,24,34) to be complete derivatives which are integrated out in
 Heavenly equations. In such a way the $q$-deformed version of the
 Heavenly equations is of the third order. As usual, when we take the
 limit $q \rightarrow 1$
 we recover the usual first and second Heavenly equations. A similar
 situation occurs for the two-dimensional Chiral Model representation of
 the Self-dual gravity.

 Though the understanding of the $q$-deformation of infinite dimensional
 Lie algebras is still
 incomplete, the  impact in physics is direct. Any advance in
 this direction will give new insights for Self-dual gravity and perhaps in
 its quantized version.

\vskip 1truecm
\begin{center}
 {\Large \bf Acknowledgements}
\end{center}
 We wish to thank Profs. M. Przanowski and Piotr Kielanowski for very useful
comments. We are indebted to Profs. I. Bakas, G.Torres del Castillo and C.
Zachos for very useful comments and suggestions.  One of us (H.G.C.) want to
thank Prof. A. Zepeda for

much encourage and help. Finally, to CONACyT and SNI for support.

\end{document}